\documentclass[twocolumn,twoside,showkeys,preprintnumbers,amsmath,amssymb,secnumarabic,nofootinbib, nobibnotes,epl,superscriptaddress]{revtex4}

\usepackage{epsfig}

\usepackage{graphicx}

\usepackage{fancyhdr}

\usepackage{pslatex}

\usepackage{subfigure}

\begin{document}



\title{A characterization of  the scientific impact of Brazilian institutions}



\author{Aristoklis D. Anastasiadis}
\affiliation{Electrical and Computer Engineering Department, University of Patras, Rio,  Achaia 26500, Greece}
\affiliation{Centro Brasileiro de Pesquisas Fisicas, Rua Xavier Sigaud 150 22290-180 Rio de Janeiro Brazil}

\author{Marcelo P. de Albuquerque}
\affiliation{Centro Brasileiro de Pesquisas Fisicas, Rua Xavier Sigaud 150 22290-180 Rio de Janeiro Brazil}

\author{Marcio P. de Albuquerque}
\email[Corresponding Author: ]{mpa@cbpf.br}
\affiliation{Centro Brasileiro de Pesquisas Fisicas, Rua Xavier Sigaud 150 22290-180 Rio de Janeiro Brazil}

\begin{abstract}
In this paper we studied the research activity of Brazilian Institutions for all sciences and also their performance in the area of physics between $1945$ and December $2008$. All the data come from the Web of Science database for this period. The analysis of the experimental data shows that, within a nonextensive thermostatistical formalism, the Tsallis \emph{q}-exponential distribution $N(c)$ can constitute a new characterization of the research impact for Brazilian Institutions.
The data examined in the present survey can be fitted successfully by applying a universal curve namely, $N(c) \propto 1/[1+(q-1)\; c/T]^{\frac{1}{q-1}}$ with $q\simeq 4/3$ for {\it all} the available citations $c$,  $T$ being an ``effective temperature''. The present analysis ultimately suggests that via the ``effective temperature'' $T$, we can provide a new performance metric for the impact level of the research activity in Brazil,
taking into account the number of the publications and their citations. This new performance metric takes into account the ``quantity'' (number of publications) and the ``quality'' (number of citations) for different Brazilian Institutions. In addition we analyzed the research performance of Brazil to show how the scientific research activity changes with time, for instance between 1945 to 1985, then during the period 1986-1990, 1991-1995, and so on until the present. Finally, this work intends to show a new methodology that can be used to analyze and compare institutions within a given country.


\end{abstract}


\maketitle

\thispagestyle{fancy}

\setcounter{page}{1}

\section{ Introduction}

The analysis of the citations of scientific papers is an important issue that can enable a better understanding of the research activity
of the authors and the institutions~\cite{Hirsch05,Times08}. The evaluation of the productivity of individual scientists has traditionally relied on the number of papers they have published. It is becoming popular to use citation analysis as a bibliometric tool for the evaluation of the scientific and academic
performance for individual researchers \cite{Hirsch05}, journals~\cite{KaterattanakulHH03,SolariM02},
universities \cite{VogelW84,Times08} even entire countries \cite{AnastasiadisMMD09}.
Nowadays, with the easy access to the Internet and to large databases,
including the Web of Science~\cite{WOS}, the comparison of the impact of scientific contributions is a much easier and more rapid process.

Research productivity is usually measured by taking into account two different variables, namely
the number of total publications and their citations. The first measure reflects research quantity and the other reflects research impact. The degree to which published works are cited by other authors is generally considered as a reflection of the quality of those works~\cite{Phelan1999}.
Prior citation works have analyzed a wide variety of factors such as the distribution of
citation rates \cite{Redner1998EPJB,LehmannLJ03,TsallisMA00}.

A stretched exponential fitting was applied for modeling citation distributions based on multiplicative processes \cite{LaherrereS98}. Lehmann \cite{LehmannLJ03} attempted to fit both a power law and stretched exponential to the
citation distribution of 281\,717 papers in the SPIRES database and showed
it is impossible to discriminate between the two models. Redner analyzed the ISI and Physical Review databases \cite{Redner1998EPJB}.
In Redner's work the applied fitting distribution had only partial success while the same numerical data for large citation count $c$ showed that it can be fitted
quite satisfactorily with a single curve by using nonextensive thermostatistical formalism~\cite{TsallisMA00}. Another fitting distribution that was applied was the lognormal distribution, which was used in order to measure the research activity \cite{RadicchiFC08}. A recent characterization of scientific impact has been conducted using Tsallis $q$-exponential distribution \cite{AnastasiadisMMD09}. In that work the scientific research activity was considered in terms of the number of publications and number of citations using data from Thomson ISI Web of Science database~\cite{WOS} for many different countries from Latin America, Europe and South Africa. That study showed that the data for all the tested countries can be satisfactorily fitted with a {\it single} curve, which naturally emerges within the Tsallis theory \cite{Tsallis99BJP}.

In this work further study has been done for the Brazilian scientific community. Traditionally, researchers and institutions have been evaluated by peer review,
which is the main mechanism for merit assessment for funding, appointment, and promotion decisions. There is also currently a global trend towards developing
and broadening the use of bibliometric indicators to help these decisions \cite{daLuzPMGCF08}.
The experimental data shows that each year there is an
increase in Brazilian contribution to international science (this is obtained by the total number of publications). The number of Brazilian authors and the number of Brazilian publications in the international scientific literature has grown substantially during the last decades \cite{WOS}. Many studies have been done to analyze the Brazilian scientific activity further and also provide a performance metric for the Brazilian Institutions \cite{MeneghiniPC08,daLuzMPMGCF08,MugainiPM08}

This manuscript provides an analysis of the scientific citations of the Brazilians institutions and their impact within a nonextensive thermostatistical formalism, the Tsallis \emph{q}-exponential distribution $N(c)$, $N(c) \propto 1/[1+(q-1)\; c/T]^{\frac{1}{q-1}}$ with $q\simeq 4/3$ for {\it all} the available citations $c$,  $T$ being an ``effective temperature''. Emphasis is also given on the performance of the Brazilian Institutions of Physics and Physics departments of Brazil's universities. The outputs of this study could be useful for the national Brazilian agencies, such as CAPES (Coordenadoria de Aperfeiçoamento de Pessoal de Nivel Superior) and other research support agencies, which are responsible for creating and assessing programs and projects. Finally, the ``effective temperature'' will be a scientific metric for the Brazilian sciences' growing performance and will help Brazilian agencies in the evaluation process of the research programs.

\section{Nonextensive Statistical Mechanics and Tsallis $q$-exponential distribution}
\label{Tsallis distribution}
Nowadays, the idea of nonextensivity has been used in many
applications. Nonextensive statistical mechanics has been applied successfully in physics (astrophysics, astronomy, cosmology,
nonlinear dynamics)~\cite{Shibata03}, biology~\cite{Upadhyaya01},
economics~\cite{Tsallis03}, human and computer sciences \cite{AnastasiadisM2004a,AnastasiadisM06,AlexandraTTMT04} and provide interesting insights into a variety of physical
systems, and among others \cite{biblio}).

Nonextensive statistical mechanics is based on Tsallis entropy. Tsallis statistics \cite{Tsallis99BJP} is currently considered useful in describing the thermostatistical properties of nonextensive systems; it is based on the generalized entropic form \cite{Tsallis88}:
\begin{equation}\label{eq:tsallis Sq}
    S_{q}\equiv k\; \frac{1-\sum_{i=1}^{W}p_{i}^{q}}{q-1} \;\;\;\;
(q \in \Re),
\end{equation}
where $W$ is the total number of microscopic configurations, whose probabilities are $\{p_i\}$, and $k$ is a conventional positive constant. When $q=1$ it reproduces the  Boltzmann-Gibbs entropic form $S_{BG}=-k\sum_{i=1}^W p_i \ln p_i$. The nonextensive entropy $S_{q}$ achieves its extreme value at the equiprobability $p_{i}=1/W,\forall i$, and this value equals $S_{q}=k \frac{W^{1-q}-1}{1-q}$ ($S_1=S_{BG}=k \ln W$)~\cite{Tsallis88,GellMannT04}.
The Tsallis entropy is nonadditive in such a way that, for statistical independent systems $A$ and $B$, the entropy satisfies the following property:
\begin{equation} \label{nonadditive}
\frac{S_q(A+B)}{k}=\frac{S_q(A)}{k} + \frac{S_q(B)}{k} + (1-q)\frac{S_q(A)}{k} \frac{S_q(B)}{k} \,.
\end{equation}
It is subadditive for $q>1$, superadditive for $q<1$, and, for $q=1$, it recovers the BG entropy, which is additive ~\cite{GellMannT04}.  The Boltzmann factor is generalized into a {\it power-law}.
The mathematical basis for Tsallis statistics includes $q$-generalized expressions for the logarithm and the exponential functions which are the $q$-logarithm and the $q$-exponential functions. The $q$-{\em exponential function}, which reduces to $exp(x)$ in the limit $q$ $\rightarrow 1$, is defined as follows
\begin{equation}\label{eq:tsallis q-exp}
    e_{q}^{x}\equiv [1+(1-q) x]^{\frac{1}{(1-q)}}=\frac{1}{[1-(q-1)x]^{\frac{1}{(q-1)}}}  \;\;\;\;(e_1^x=e^x)  \, .
\end{equation}
We remind that extremizing entropy $S_{q}$  under appropriate constraints we obtain a probability distribution, which is proportional to $q$-{\em exponential function}.

In this work, we focus on the analysis of the distribution of citations of scientific publications, more precisely those that have been catalogued by the Institute for Scientific Information (ISI) for the Brazilian Institutions and for the whole of Brazil.
The proposed fitting distributions follow from the nonextensive formalism as  $N(x) \propto 1/[1+(q-1)\; c/T]^{\frac{1}{q-1}}$.
In this study we adopt the following expression:
\begin{equation}
N(c) = N(2)\, \exp_{q}^{-\frac{c-2}{T}}
\end{equation}
where $N(2)$ is the number of papers with two citations, and, as already mentioned, $T$ plays the role of an effective temperature.

\section{Thomson ISI Web of Knowledge- Data Acquisition}
Traditionally, the most commonly used source of bibliometric data is Thomson ISI Web of
Knowledge, in particular the (Social) Science Citation Index and the Journal Citation Reports
(JCR), which provide the yearly Journal Impact Factors (JIF) \cite{WOS}.
The subject categories and terminology provided by ISI are widely recognized by many researchers and
scientometricians in their studies and are relatively simple to use \cite{AnastasiadisMMD09,RadicchiFC08}.
The Institute for Scientific
Information has made an industry of providing citation data to libraries
since the mid-1960s; the products are currently available as part of Thomson/ISI. Although the ISI database
has a few shortcomings, overall it gives a wide coverage of most research fields \cite{May97}. Therefore in our survey
we utilize Thomson/ISI Web of Science database to study the distribution of the citation within a variety of countries.

To obtain all the necessary data we developed a program which automatically downloads the ISI
bibliographic information.
We take into account all the document types, e.g. articles, proceeding papers, meeting abstracts, etc, for all the
available subject areas, for instance neurosciences, mathematics, chemistry
etc, to select all the data for the Brazil and then the same procedure for the Brazilian Institutions and the departments
and institutes of physics that we are interested in.

The program was written in Delphi 7 and uses the TWebBrowser component.
This component provides access to the Web browser functionality and saves all
the ``html'' pages. When the page was completely downloaded, an
OnDownloadComplet event was generated and we went automatically to the next
``html'' page. When all pages were downloaded we processed each ``html'' page to
obtain the specific information that we were interested in using the TPerlRegEx
component from the open source PCRE library [http://www.pcre.org/]. In this
case, we gathered the number of citations for each publication and the total number of the published papers, for each
Institution. We applied filters to take all these data sorted by the times cited,
using the Citation databases namely Science Citation Index Expanded
($SCI-EXPANDED$ 1945--present), Social Sciences Citation Index ($SSCI$
1956--present, and Arts and humanities Citation Index ($AandHCI$
1975--present). All the data was captured during December 2008.

\section{Presentation of Results}
\begin{figure}
\includegraphics[width=0.52\textwidth]{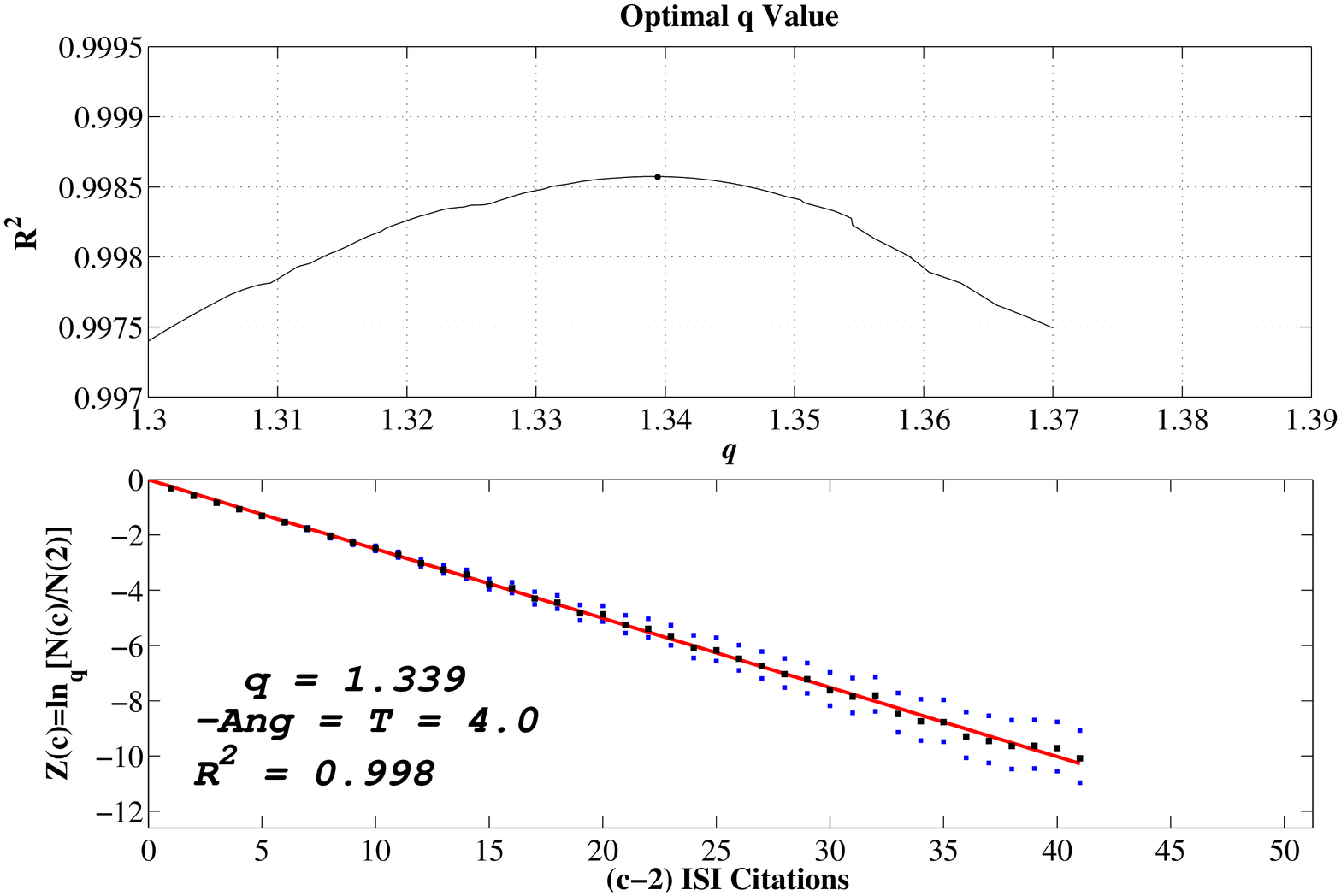}
\includegraphics[width=0.5\textwidth]{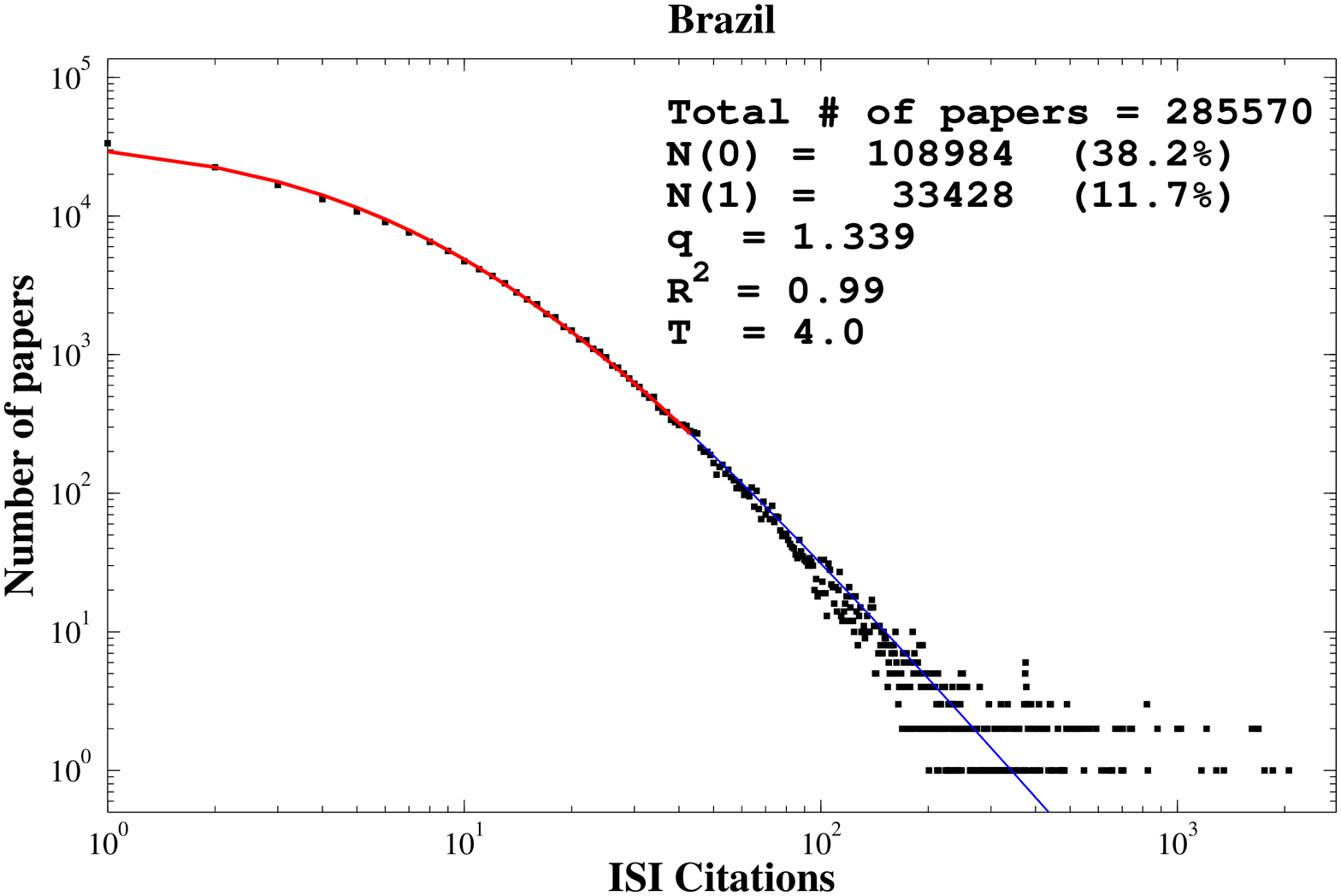}
\caption{Process to obtain best $q$ and $T$ (up). \\
Probability distribution for citations of Brazil country (down)}
\label{Brazil-linear-log-fitting}       
\end{figure}

Firstly, we are going to present the data for the whole of Brazil captured until December 2008 and then describe the procedure that we follow to conduct the final citations fitting. All the papers included in the Web of Science and having at least one author with at least one affiliation address in Brazil have been collected. This means that the work includes all the documents with at least one Brazilian address with citations till December 2008.
Research done by Brazilians abroad, i.e with only foreign addresses, is disregarded in the considered database. Note that the data and results are presented on a log-log scale.
Initially we evaluate the values of $q$ in order to find its optimal value, and then, with this value, we move to the final fitting in order to determine $T$.
The corresponding angle gives the optimal value of the effective temperature $T$ (Figure~\ref{Brazil-linear-log-fitting}).
With these two values ($q$ and $T$) we present the fitting in a log-log diagram. In the Brazil case a remarkably good fitting can be done with  $q=1.339$ and $T=4.0$. This temperature provides good evidence about the impact of the published papers, and enables a ranking. Figure~\ref{Brazil-linear-log-fitting} illustrates the entire process.

Next we investigate how the temperature changes during the years. As the temperature is a characterization of the scientific impact its evolution over
the years can offer a deeper understanding of how the Brazilian research activity evolved.  Figure 2 presents the temperature for each period that we study, for instance between 1945 to 1985, then during the period 1986-1990, 1991- 1995 and so on. This histogram highlights how the scientific research activity changes with time. It is remarkable how effective temperature is as a reliable performance metric for the research activity of Brazil. This part of the analysis uses the entire available-year publication window for all disciplines for papers published between 1945 to December 2008. Note that for the last periods from 2001 to 2004 and 2005 to 2008 there has not been enough time for the publications to become widely known to the scientific community so the number of their citations is small. Thus the overall temperature is smaller as there is this delay. Also Figure 2 (right) illustrates the performance of Brazil in Physics domain. 39\,617 papers (8\,688 zero citations, $(21.9\%)$) are published in Physics until January 2009 giving T=4.44, which characterizes the overall research performance of our tested Brazilian society of Physics.

\begin{figure*}[ht!]
\begin{center}
\includegraphics*[width=8.9cm]{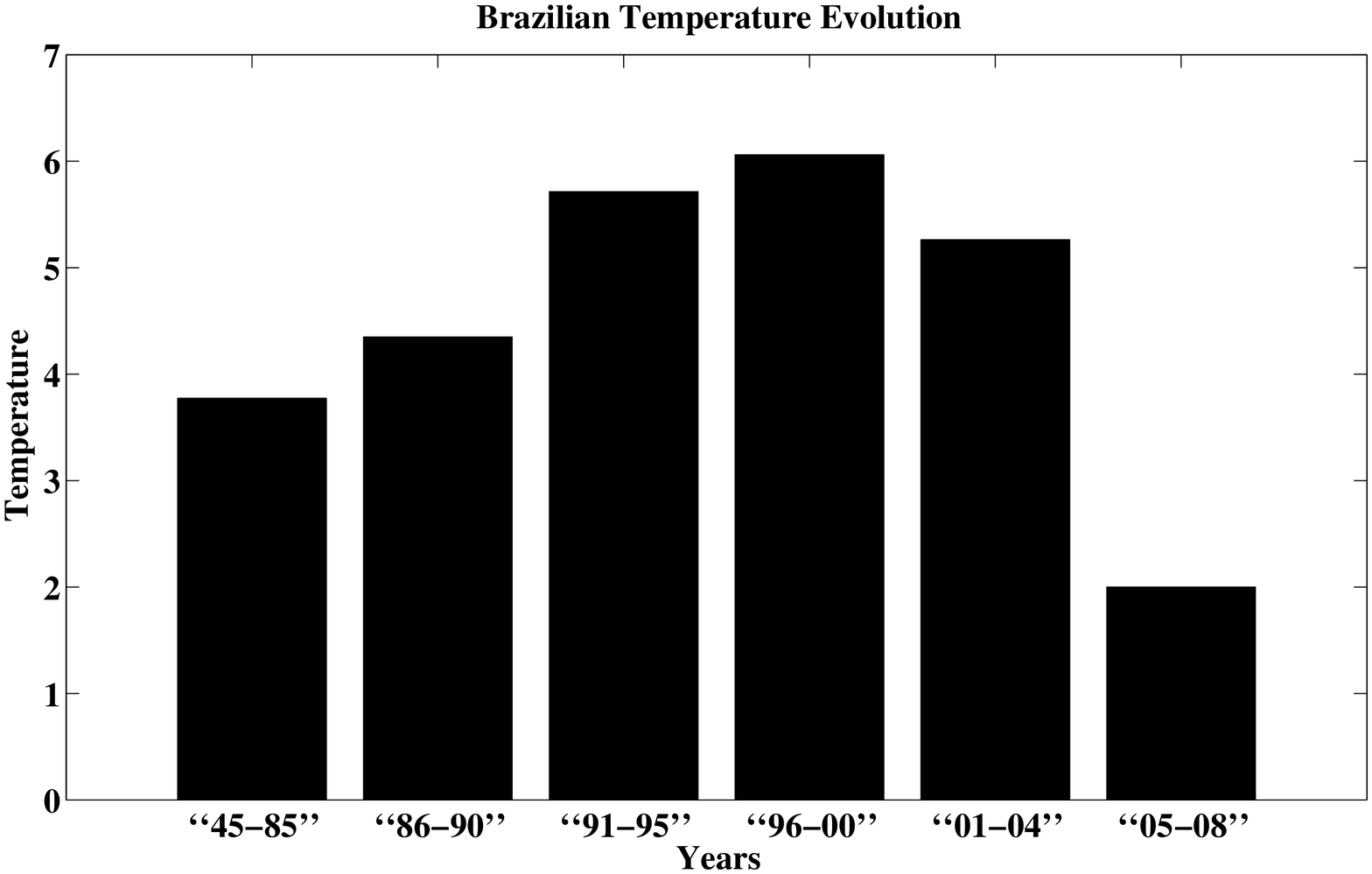}
\includegraphics*[width=8.9cm]{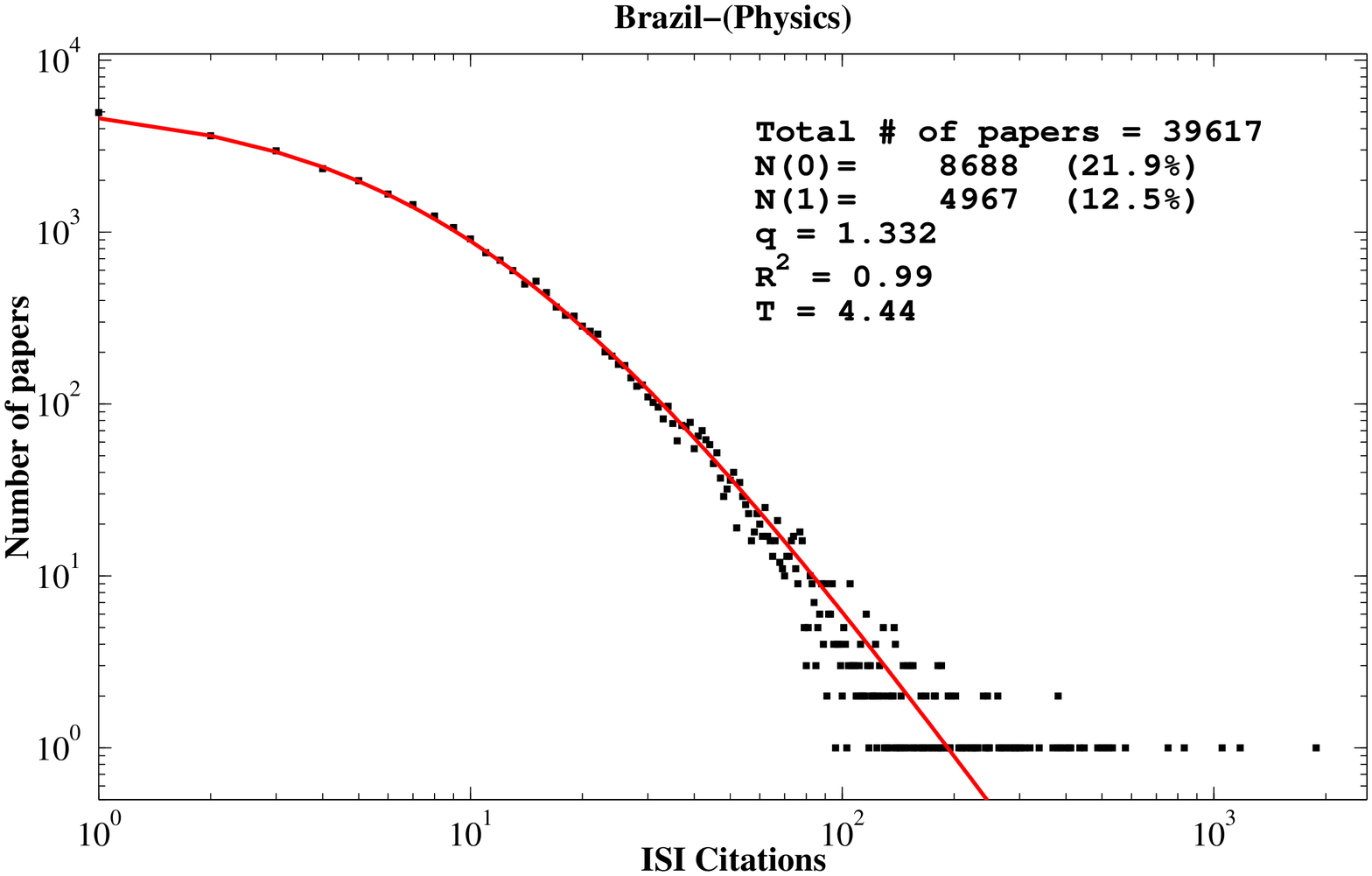}
\caption{Brazilian evolution of the effective temperature within the tested period (left) and the Probability distribution for citations of Brazil in Physics (right)}
\end{center}
\label{fig:Brazilian temperature Evolution and Brazil Physics}       
\end{figure*}

Note that the results for ``Brazil'' do not represent the average of the particular Brazilian institutions that we are considering in the Tables but all the Brazilian institutions. This happens because these results are taken by placing ``Brazil'' in the address field. It should also be clear that when we refer to ``Brazil Physics'', it is the average research performance for all the Brazilian institutions in the area of physics and not only the tested Brazilian institutes, i.e in this case  we apply the word ``Brazil'', and Physics (``Fis'') in the address field to obtain these results. Finally, in the tables \ref{table:Ranking Institutions} and \ref{table:Ranking Institutions on Physics} we study the institutions with temperature greater or equal to the whole Brazilian temperature, i.e $T\geq4.0$ .

Table \ref{table:Total Brazilian Institutions Papers} presents the total number publications, and the percentage of zero,  and once cited papers
for the tested Brazilian Institutions. University of Sao Paulo (USP) achieves the highest publication productivity with 66\,404 published papers.
Then University Estadual Campinas (UNICAMP) and Federal University of Rio de Janeiro (UFRJ) publish 24\,209 and 21\,656 research papers respectively. The rest of
the tested Brazilian Institutions attain a significantly lower rate of published papers, i.e Federal University of Pernambuco (UFPE),
Federal University of Rio Grande do Sul (UFRGS), and Federal University of Fluminense (UFF) have published 6\,032 , 5\,540, 5\,318 papers respectively. Finally, the Federal University of Minas Gerais (UFMG) have 1\,887 publications.

\begin{table}
\vspace*{0.05cm}
\begin{center}
\begin{scriptsize}
\caption{Number of total publications, and the percentage of zero,  and one cited papers for the tested Institutions} \label{table:Total Brazilian Institutions Papers}
\newcommand{\m}{\hphantom{$-$}}
\newcommand{\cc}[1]{\multicolumn{1}{c}{#1}}
\renewcommand{\tabcolsep}{0.7pc} 
\renewcommand{\arraystretch}{1.0} 
\begin{tabular}{@{}lllllllll}
\hline
                &  \textbf{Total \# Papers}         & $\;\;$\textbf{\# Zero citations}        &\textbf{\# One citations}      \\
Institutions     & $\;$$\sum_{c=0}^{\infty} N(c)$    & $\;\;$$N(0)$ $\;\;\;\;\;\;\;\;$ $(\%)$   &$\;\;$$N(1)$ $\;\;\;\;$ $\;\;$ $(\%)$  \\
\hline
USP           &$\;\;\;$66\,404      &$\;\;\;\;\;$24\,197     $\;\;\;$(36.4\%)   &$\;\;$7\,041       $\;$($\;$10.6\%) \\
UNICAMP       &$\;\;\;$24\,209      &$\;\;\;\;\;\;\;$8\,215  $\;\;\;$(33.9\%)    &$\;\;$2\,771      $\;\;$(11.5\%)       \\
UFRJ          &$\;\;\;$21\,656      &$\;\;\;\;\;\;\;$7\,498  $\;\;$  (34.6\%)    &$\;\;$2\,591      $\;\;$(12.0\%)        \\
UFPE          &$\;\;\;\;$6\,032     &$\;\;\;\;\;\;\;$2\,067  $\;\;$ (34.3\%)     &$\;\;\;\;\;$794      $\;\;$(13.2\%)        \\
UFRGS         &$\;\;\;\;$5\,540     &$\;\;\;\;\;\;\;$2\,868  $\;\;\;$(51.8\%)    &$\;\;\;\;\;$695     $\;\;$(12.5\%)         \\
UFF           &$\;\;\;\;$5\,318     &$\;\;\;\;\;\;\;$1\,919  $\;\;$ (36.1\%)     &$\;\;\;\;\;$668      $\;\;$(12.6\%)        \\
UFMG          &$\;\;\;\;$1\,887     &$\;\;\;\;\;\;\;\;\;$ 680  $\;\;\;$(36.0\%)    &$\;\;\;\;\;$286       $\;\;$(15.2\%)          \\
Brazil        &285\,570             &$\;\;\;$108\,984           $\;\;$ (38.2\%)   &$$33\,428   $\;\;$(11.7\%)        \\
\hline
\end{tabular}
\end{scriptsize}
\end{center}
\vspace*{0.0cm}
\end{table}

Next, Table \ref{table:Ranking Institutions} presents the Brazilian Institutions in the ranking based on the temperature that we obtain through the nonextensive distribution fitting.
Notice that this ranking differs from the one presented in Table ~\ref{table:Total Brazilian Institutions Papers},
where the total amount of the published papers (quantity ranking) is shown. The effective temperature $T$ characterizes the scientific impact of the tested Institutions.
As we can perceive from Table \ref{table:Ranking Institutions}, in almost all cases the range value of the entropic index \emph{q} is around $q=4/3$.
The linear regression coefficient $R^2$ is also indicated in each case.
As we can see comparing Tables \ref{table:Total Brazilian Institutions Papers} and \ref{table:Ranking Institutions}, the rankings are quite different. Let us check UFRJ, for instance.
Although it has a relatively smaller number of papers compared to UNICAMP, its effective temperature is higher $T=4.55$.


\begin{table*}
\vspace*{0.05cm}
\begin{center}
\begin{small}
\caption{Best fitting values of \emph{q} and effective temperature\emph{T}. Note that tested Institutions are ranked according to \emph{T}} \label{table:Ranking Institutions}
\newcommand{\m}{\hphantom{$-$}}
\newcommand{\cc}[1]{\multicolumn{1}{c}{#1}}
\renewcommand{\tabcolsep}{0.6pc} 
\renewcommand{\arraystretch}{1.0} 
\begin{tabular}{@{}lllllllllll}
\hline
&   \textbf{Entropic index}           & \textbf{Linear regression}    & \textbf{Temperature}    \\
Institutions             & $q$                 & {\bf coefficient} $R^{2}$                      & \textbf{$T$}           \\
\hline
USP                         & 1.339                             & 0.99                      & \textbf{4.75}             \\
UFRJ                        & 1.300                             & 0.99                      & \textbf{4.55}              \\
UNICAMP                     & 1.330                             & 0.99                      & \textbf{4.35}           \\
UFPE                        & 1.336                             & 0.99                      & \textbf{4.08}                \\
UFF                         & 1.335                             & 0.99                      & \textbf{4.00}            \\
Brazil                        & 1.339                             & 0.99                      & \textbf{4.00}              \\
\hline
\end{tabular}
\end{small}
\end{center}
\vspace*{0.0cm}
\end{table*}
Table \ref{table:Ranking Institutions on Physics} presents the best fitting values of \emph{q} and the effective temperature \emph{T}, which characterizes
the research impact of the Brazilian Institutions with emphasis on Physics. In this analysis UFMG was not included as the available publications in the Web of Science database \cite{WOS} are few (not enough to have a good statistical analysis).
In this survey the Centro Brasileiro de Pesquisas Fisicas (CBPF) is also included .
It becomes evident from the Table  \ref{table:Ranking Institutions on Physics} that CBPF, USP and UNICAMP achieved the highest temperature in research activity in Physics by applying the new metric ($T$).
It is also worth mentioning that the Institutes/Departments of Physics of the Universities have the responsibility of both undergraduate/graduate students and are administratively located at the Ministry of Education, whereas CBPF has the responsibility of only graduate students and is administratively located at the Ministry of Science and Technology. This is possibly one of the reasons that can help this institute to achieve higher temperature.
Moreover it is important to mention at this point the performance of the UFPE and UFRJ. While the UFPE increases significantly its Temperature  on the domain of Physics the UFRJ has lower $T=4.10$ compared to the overall research impact value ($T=4.55$) in all sciences.

\begin{figure*}[ht!]
\begin{center}
\includegraphics[width=8.9cm]{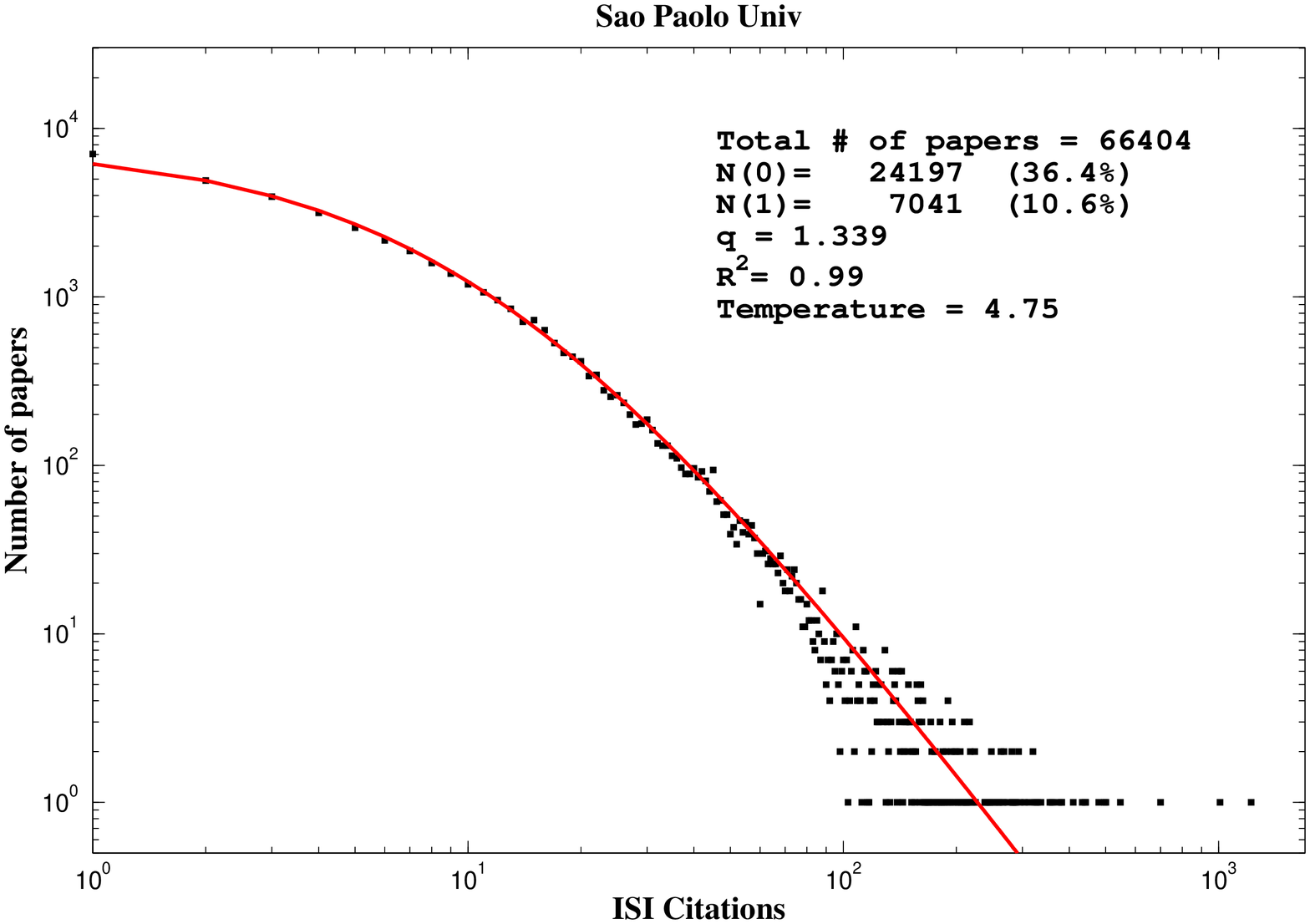}
\includegraphics[width=8.9cm]{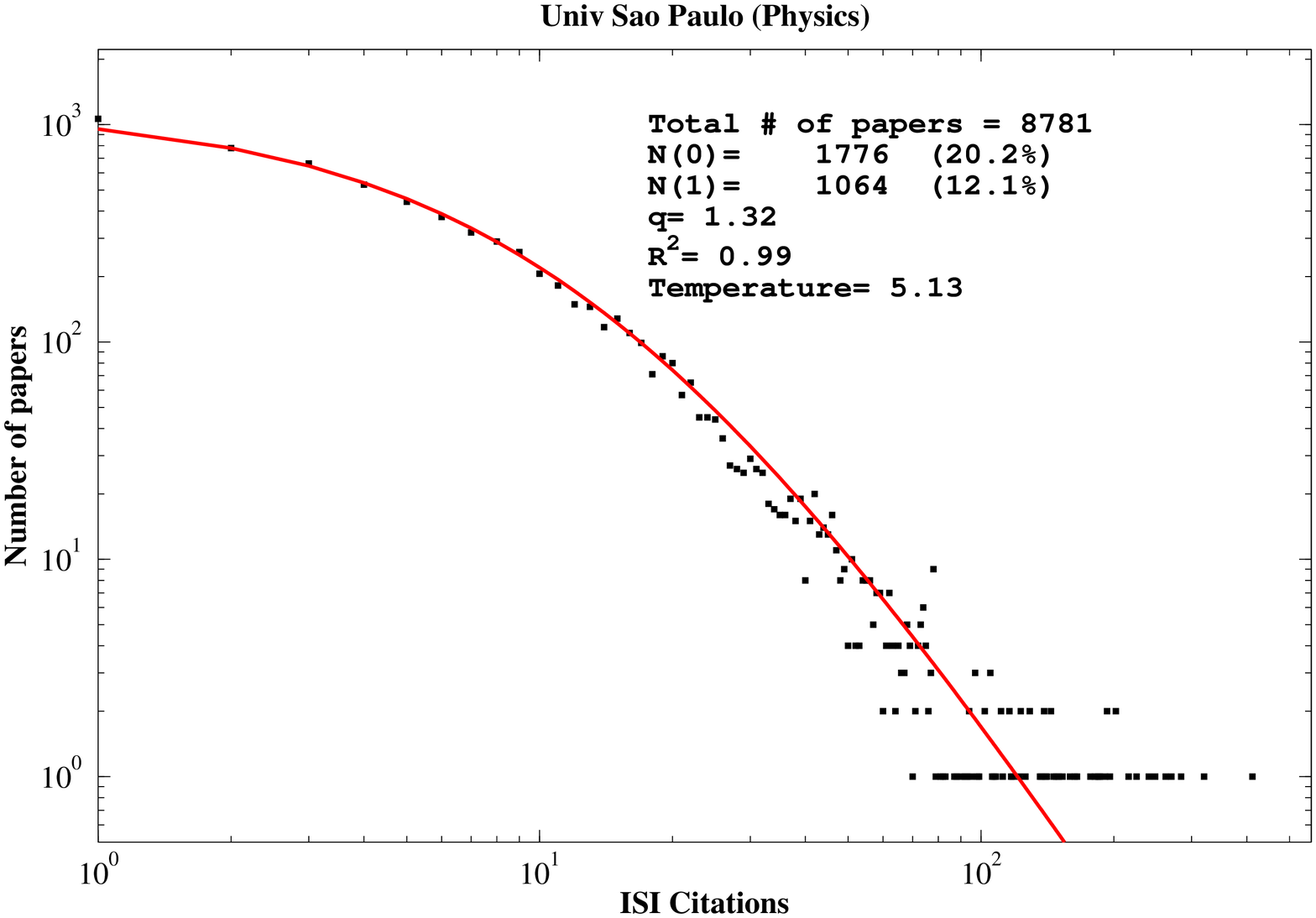}
\includegraphics[width=8.9cm]{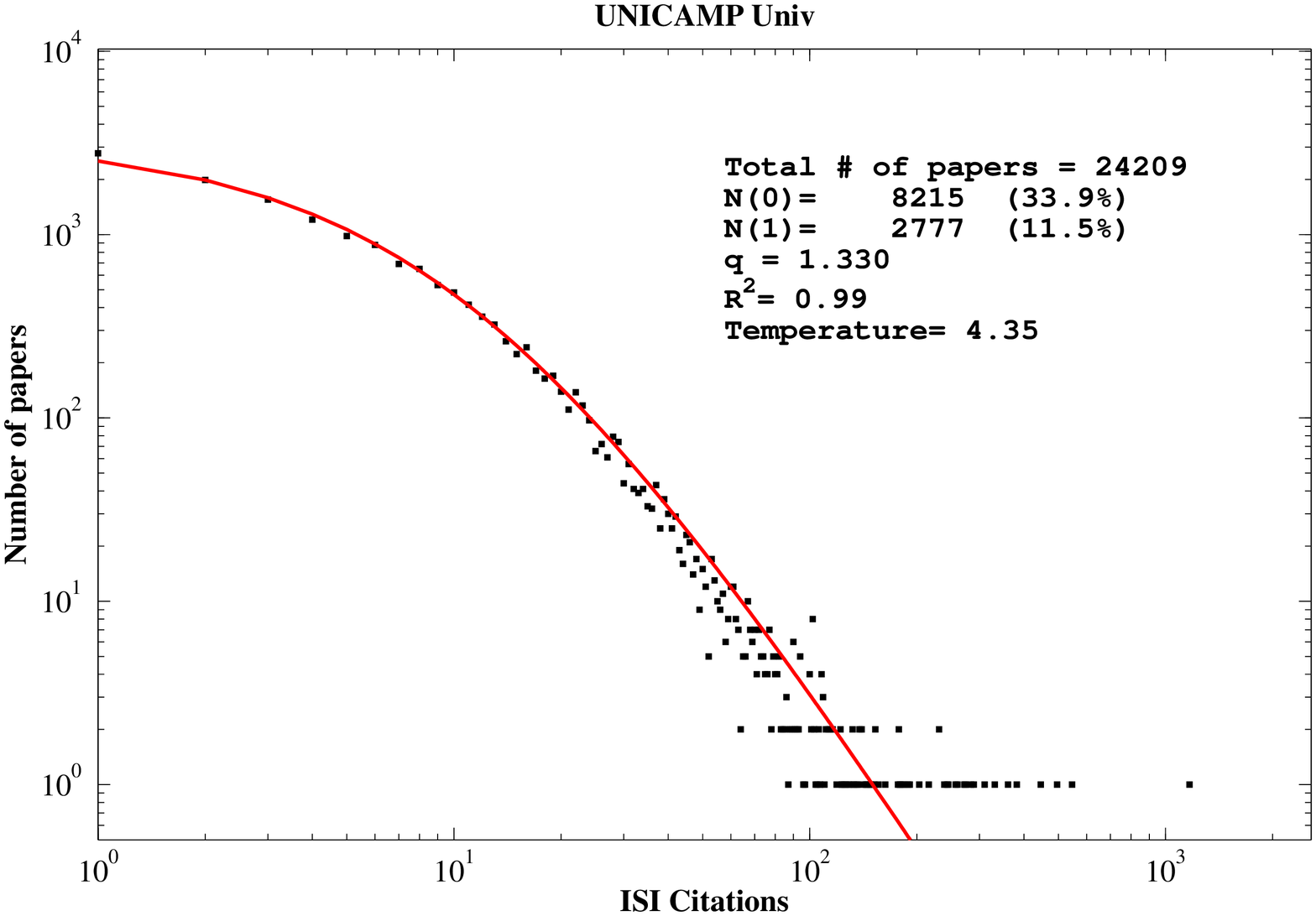}
\includegraphics[width=8.9cm]{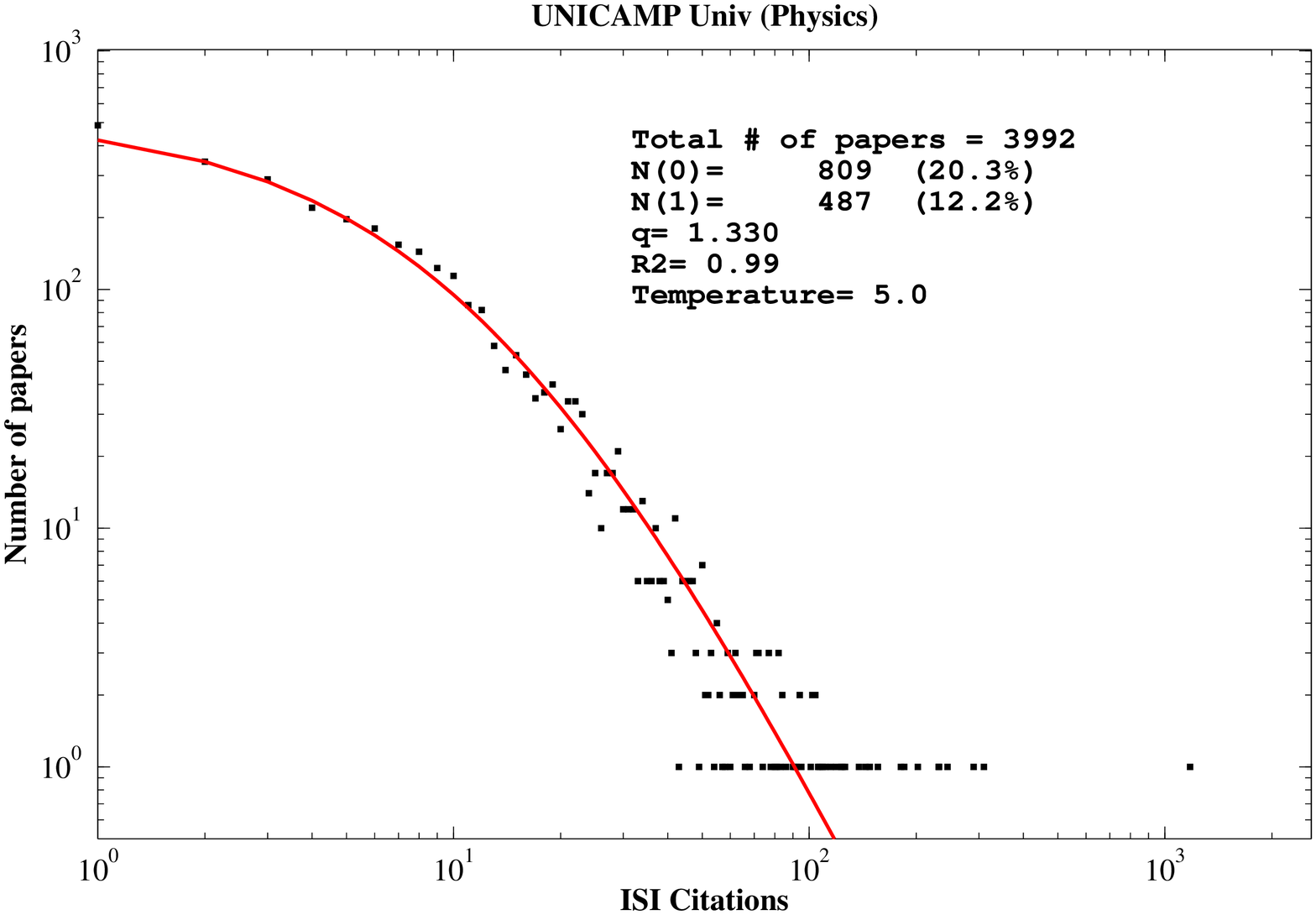}
\includegraphics[width=8.9cm]{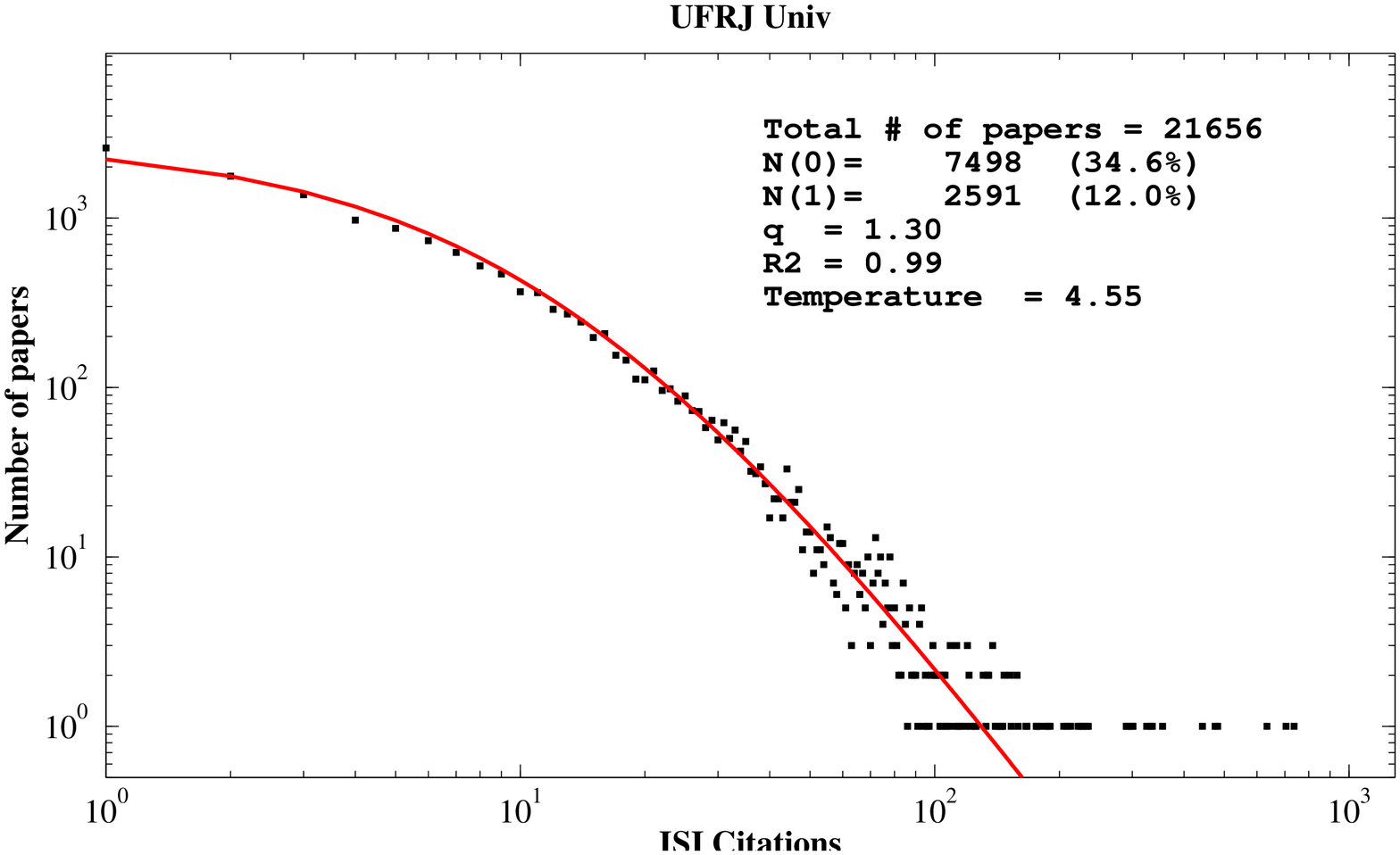}
\includegraphics[width=8.9cm]{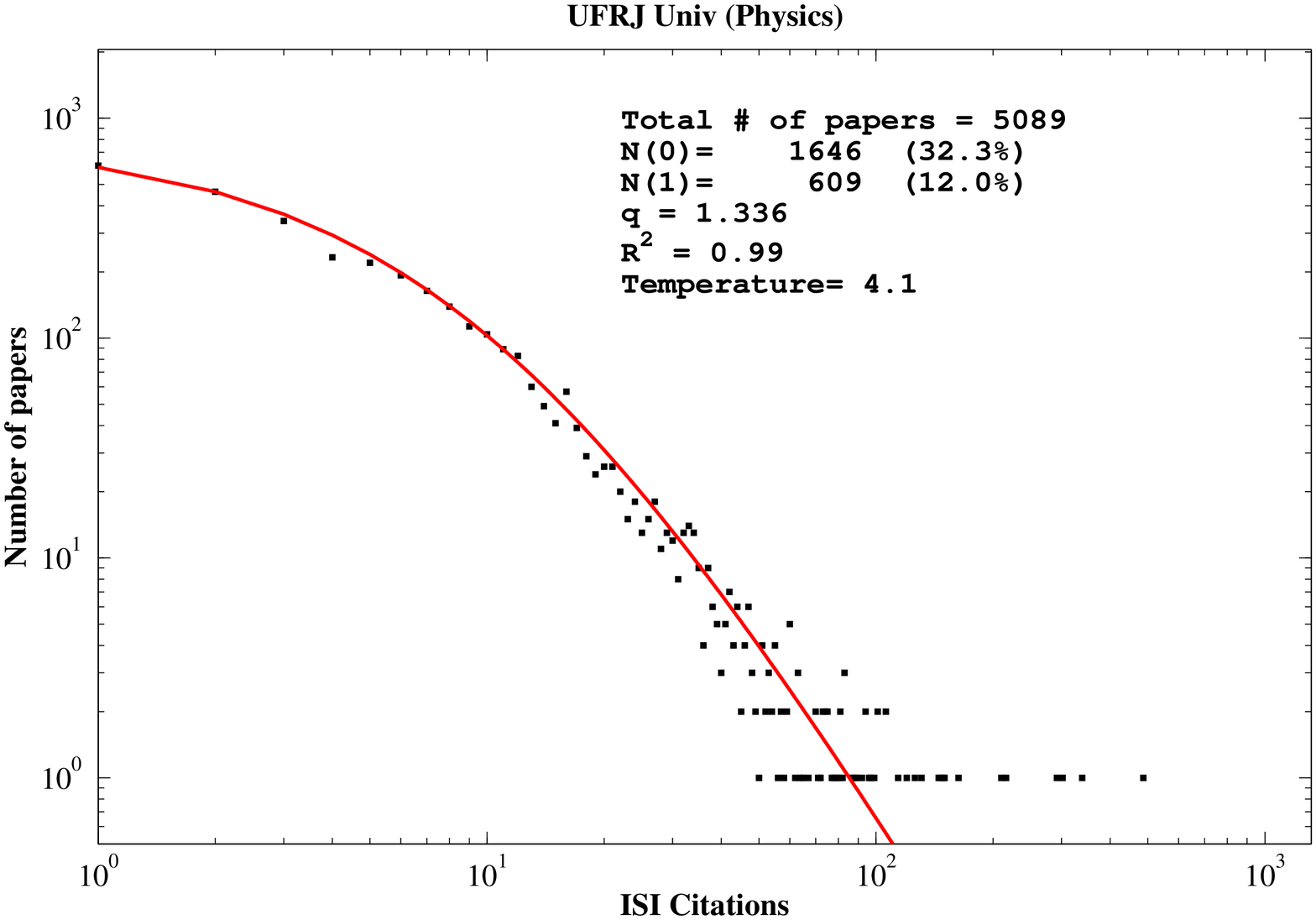}
\caption{Probability distribution for citations of Brazilian Institutions and their Physics departments}
\label{fig:Brazil Inst All sciences and Physics curves}
\end{center}
\end{figure*}

\begin{figure*}[ht!]
\begin{center}
\includegraphics[width=8.9cm]{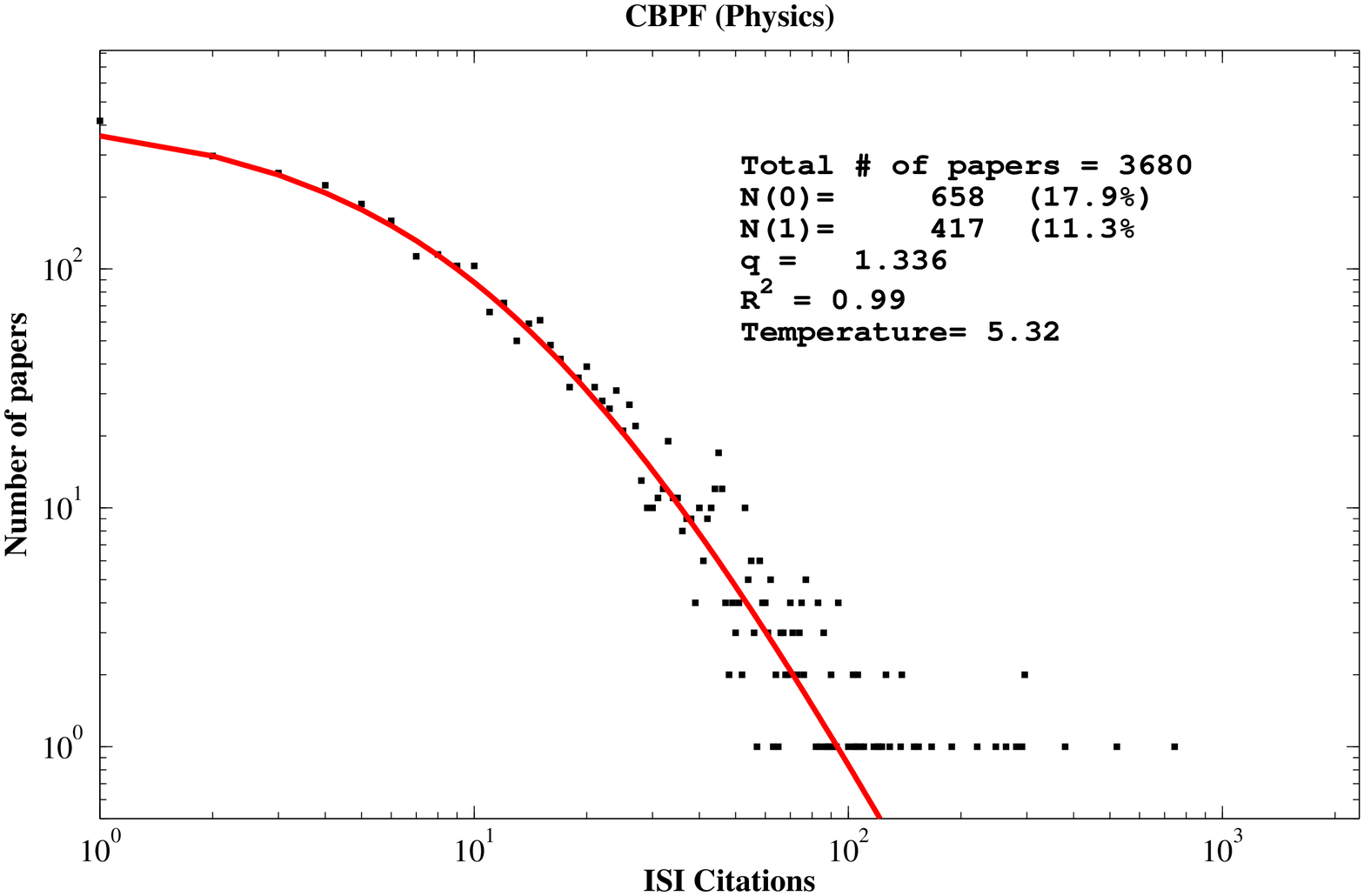}
\includegraphics[width=8.9cm]{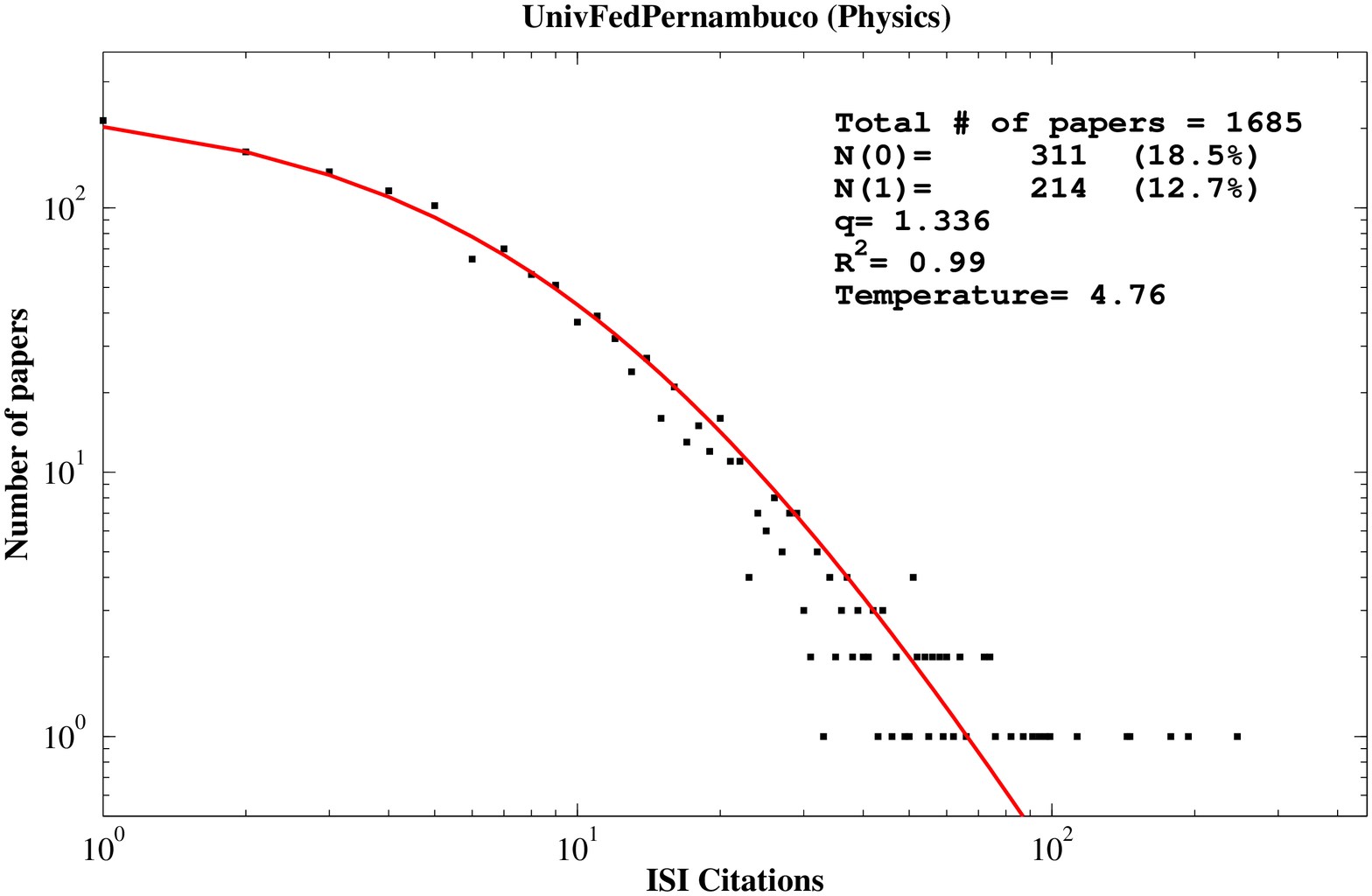}
\caption{Probability distribution for citations of CBPF (left) and UFPE (right) up to January 2009}
\label{fig:CBPF and UFPE curves}
\end{center}
\end{figure*}

\begin{table*}
\vspace*{0.05cm}
\begin{center}
\begin{small}
\caption{Best fitting values of \emph{q} and effective temperature\emph{T}. Note that tested Institutions are ranked according to \emph{T}} \label{table:Ranking Institutions on Physics}
\newcommand{\m}{\hphantom{$-$}}
\newcommand{\cc}[1]{\multicolumn{1}{c}{#1}}
\renewcommand{\tabcolsep}{0.6pc} 
\renewcommand{\arraystretch}{1.0} 
\begin{tabular}{@{}lllllllllll}
\hline
&  \textbf{Total \# Papers}               &  \textbf{\# Zero citations}   &   \textbf{Entropic index}   & \textbf{Linear regression}    & \textbf{Temperature} \\
Physics  & $\;$$\sum_{c=0}^{\infty} N(c)$          &$N(0)$ $\;\;$ $(\%)$      &$q$                         & {\bf coefficient} $R^{2}$     & \textbf{$T$}    \\
\hline
CBPF         &  3\,680                      &  $\;\;\;$ 658  $(17.9\%)$     & 1.336                             & 0.99              & \textbf{5.32}       \\
USP          & 8\,781                       &  1\,776       $$ $(20.2\%)$          & 1.320                         & 0.99                      & \textbf{5.13}      \\
UNICAMP      & 3\,992                       &  $\;\;\;$ 809  $(20.3\%)$      & 1.330                             & 0.99                      & \textbf{5.0}   \\
UFPE         & 1\,685                       &  $\;\;\;$ 311  $(18.5\%)$     & 1.336                             & 0.99                      & \textbf{4.76}    \\
UFRJ         & 5\,089                       &  1\,646  $$ $(32.3\%)$           & 1.336                             & 0.99                      & \textbf{4.10}       \\
UFF          & 1\,512                       &  $\;\;\;$ 309  $(20.4\%)$       &1.332                             & 0.99                      & \textbf{4.08}       \\
Brazil Physics      & 39\,617                      &  8\,688  $$ $(21.9\%)$          & 1.332                             & 0.99                    & \textbf{4.44}         \\
\hline
\end{tabular}
\end{small}
\end{center}
\vspace*{0.0cm}
\end{table*}
Figures \ref{fig:Brazil Inst All sciences and Physics curves} and \ref{fig:CBPF and UFPE curves} illustrate the fitting of different Brazilian Institutions using
the nonextensive distribution $N(c)$. Figure \ref{fig:Brazil Inst All sciences and Physics curves} left side shows publications of all sciences and right side demonstrates the research activity in physics domain. As we can observe the general tendency for physics science have a higher research impact than the overall university activity. Finally, Figure \ref{fig:CBPF and UFPE curves} presents the CBPF and UFPE fitting curves by applying the new characterization of citations impact. CBPF achieves the highest performance with T=5.32 and $q$=1.336. UFPE physics domain attains T=4.76 while the whole UFPE's university citations impact metric is $4.08$.

From all the above experimental results, we obtain a value of $q$ close to $4/3$.


\section{Conclusions}
Nowadays the number of citations is among the most widely used measures of academic performance. Extended study of citation distributions helps to understand better the mechanics behind citations and can objectively establish a comparative measure for scientific performance. Citations of scientific papers constitute in fact a connection network consisting of authors (nodes) and directed links (citations) among them. Recently, connection networks have been described, studied, characterized and represented by parameters using typical concepts in the area of Complex Systems.

The entropic index $q$ in Tsallis entropy is usually interpreted as a quantity characterizing the degree of nonextensivity of a system. An appropriate choice of the entropic index $q$ to nonextensive physical systems still remains an open field of study.
In some cases, the physical meaning of the index $q$ is unknown; it provides nevertheless
new possibilities of comparison between theoretical approaches and experimental data. Other cases are better understood, and then $q$ has a clear
physical meaning, either at a microscopic or at a mesoscopic level, or both.

In this paper we characterize the citations impact of the Brazilian institutions using the Tsallis q-exponential distribution. We also show how the scientific research activity changes with time, between six periods from 1945 to 2008. The present study provides a new performance metric based on Nonextensive Statistical Mechanics for ranking and evaluating institutions' research production. The proposed Tsallis q-exponential distribution satisfactorily describes Institute of Scientific Information citations for Brazilian institutions and Brazilian physics departments between 1945 and December 2008.

\begin{figure*}[ht!]
\begin{center}
\includegraphics[width=14cm]{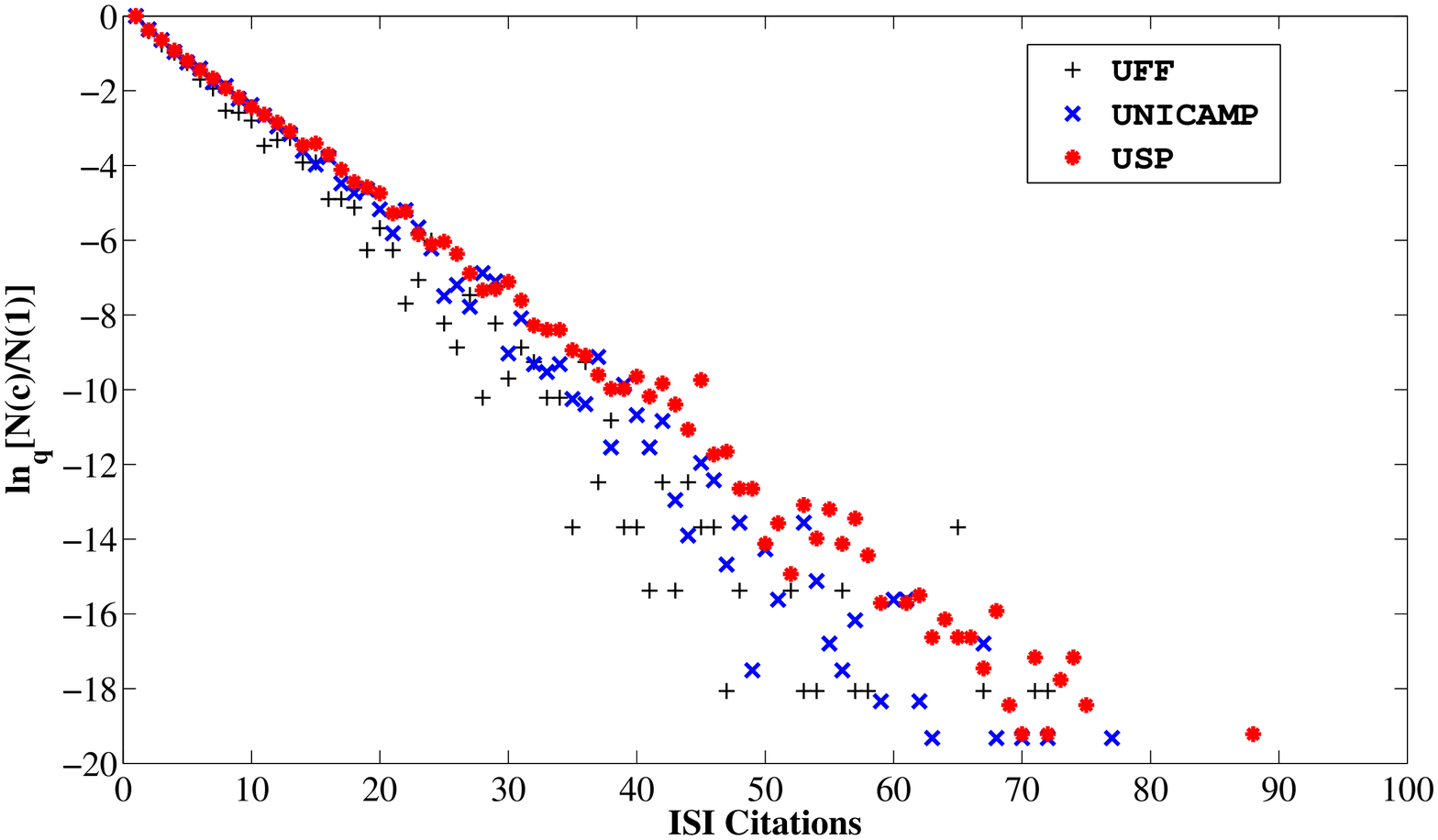}
\caption{$\ln_{q}[{N(c)/N(1)}]$ versus $(c-1)$ for 3 Brazilian Universities (UFF, UNICAMP, USP).
The figure is a zoom with an upper limit of 100 citations.}
\label{fig:UFF UNICAMP USP data curves}
\end{center}
\end{figure*}

Our study provides evidence that the citation distribution for all tested cases within this period could be the Tsallis q-exponential distribution. Our findings in this work gives an evidence for the effectiveness of $T$, and the ranking that we proposed based on the Temperature. Figure \ref{fig:UFF UNICAMP USP data curves} illustrates the $q$-logarithmic number of publications $\ln_{q}[{N(c)/N(1)}]$  versus the $(c-1)$ number of citations for three different Brazilian universities (UFF, UNICAMP, USP). USP has the higher citation impact, the UNICAMP an intermediate  $T$ and UFF lower temperature than the average. It is important to notice that $(-1/T)$ corresponds to the average slope associated with each university. It also gives an explanation for the meaning of $T$, and the ranking that we proposed based on the new performance metric $T$.

It is remarkable how the proposed nonextensive distribution fits satisfactorily all cited papers for all the institutions. This part of the analysis uses the entire available-year publication window for all disciplines for papers published between 1945 to December 2008. The present article also focuses on the performance of the Brazilian Institutions and their activities in physics science. In the present study we used a single database for the extraction of the articles, and their number of citations. The ISI/Web of Science was chosen because it is one of the main databases providing information on citations. Although our strategy might have left publications out of the analysis, we believe that the sample of articles was representative of the core international scientific production of the Brazilian Institutions. The new performance metric of citations impact is a balanced combination of ``quantity'' (number of publications) and ``quality'' (number of citations). These are the main factors of this performance metric. Keeping in mind that citation rate reflects the use and impact of scientific information, not necessarily expressing quality.

This work intends to show how the new methodology can be used to analyze and compare institutions within a given country. A case study of certain Brazilian institutions and their physics departments is used to investigate the effectiveness of the new characterization of citations impact. Future work can address other scientific fields in these important Brazilian universities or universities of other countries and how they evolved observing the same analyzed period of time. It is also important to study cases of universities, countries or other scientific institutions with extremely high number of zero or one citations and observe the impact of their research activity. The extent to which this number of citations affects the proposed performance metric will be a field of further study.

\bigskip

\noindent
{\bf Acknowledgements}
The authors thank Professor C.Tsallis for helpful discussions, D. B. Mussi for the program development, the support from the National Council for Scientific and Technological Development (CNPq) of the Brazilian Ministry of Science and Technology,
the State of Rio de Janeiro Research Foundation (FAPERJ) and the Brazilian Coordination Office for the Improvement of Staff with Higher Education (CAPES).


\begin{thebibliography}{99}

\bibitem{Hirsch05}{J.E. Hirsch, An index to quantify an individual's scientific research output, {\em Proc. Nat. Acad.
Science}, {\bf 102}, 165--169, (2005).}

\bibitem{Times08}{TIMES higher education World University Rankings 2008: http://www.timeshighereducation.co.uk/}

\bibitem{WOS}{ISI Web Of Knowledge, http://portal.isiknowledge.com}

\bibitem{KaterattanakulHH03}{P. Katerattanakul, P. Han, and B. Hong, Objective quality ranking of computing journals,
{\em Communications of the ACM}, {\bf 46} no 10, 111–-114, (2003).}

\bibitem{SolariM02}{A. Solari., M. Magri, A new approach to the SCI Journal Citations Reports, a system for
evaluating scientific journals. {\em Scientometrics}, {\bf 47} no. 3, 605–-625, (2002).}

\bibitem{VogelW84}{D.R. Vogel, J.C. Wetherbe, MIS research: a profile of leading journals and universities. The
Database for Advances in Information Systems, {\bf 16} no. 1, 3-–14, (1984).}

\bibitem{AnastasiadisMMD09}{A.D. Anastasiadis, M.P de Albuquerque, M.P de Albuquerque and D.B. Mussi,
Tsallis $q$-exponential describes the distribution of scientific citations - A new characterization of the impact,
arXiv:0812.4296v1, 2009}

\bibitem{AnastasiadisM2004a}{A.D. Anastasiadis, and G.D. Magoulas.,
Nonextensive statistical mechanics for hybrid learning of neural
networks, {\em Physica A: Statistical Mechanics and its Applications}, {\bf 344}, pp. 372--382, (2004).}

\bibitem{AnastasiadisM06}{A.D. Anastasiadis, and G.D. Magoulas.,
Evolving Stochastic Learning Algorithm based on Tsallis Entropic index, {\em The European Physical Journal B}, {\bf 50}, 277-–283, (2006). }

\bibitem{RadicchiFC08}{F. Radicchi, S. Fortunato and C. Castellano.,
Universality of citation distributions: towards an objective measure of scientific impact,
Proc. Natl. Acad. Sci. USA 105, 17268--17272, (2008)}

\bibitem{daLuzPMGCF08}{M.P. da Luz, C.M. Portella, M. Mendlowicz, S. Gleiser, E.S.F Coutinho and I. Figueira,
Institutional h-index: The performance of a new metric in the evaluation of
Brazilian Psychiatric Post-graduation Programs, Scientometrics, {\bf 77} no. 2, 361--368, (2008).}

\bibitem{Phelan1999}{T.J. Phelan, A compendium of issues for citation analysis., {\em Scientometrics}, {\bf 45} no. 1, 117–-136, (1999).}

\bibitem{Redner1998EPJB}{S. Redner, How popular is your paper? An empirical study of the citation
distribution, {\em Eur. Phys. J. B}, {\bf 4}, 131--134, (1998).}

\bibitem{LaherrereS98}{J. Laherrere. and D. Sornette,  Stretched exponential distributions in nature and
economy: ``fat tails'' with characteristic scales. {\em The European Physical Journal B -
Condensed Matter}, {\bf 2} no. 4, pp. 525–-539, (1998).}

\bibitem{LehmannLJ03}{S. Lehmann, B. Lautrup, and A.D. Jackson, Citation networks in high energy
physics. {\em Physical Review E} (Statistical, Nonlinear, and Soft Matter Physics),
{\bf 68} no. 2, (2003).}

\bibitem{MeneghiniPC08}{R. Meneghini, Abel L. Packer, L.N. Calo, Articles by Latin American Authors in Prestigious
Journals Have Fewer Citations, {\em Plos one}, (2008)}

\bibitem{MugainiPM08}{R. Mugnaini, A.L. Packer, and R. Meneghini, Comparison of scientists of the Brazilian
Academy of Sciences and of the National Academy of Sciences of the USA on the basis of the h-index,
{\em Brazilian Journal of Medical and Biological Research} {\bf 41}, 258--262, (2008)}

\bibitem{daLuzMPMGCF08}{M.P da Luz, C. Marques-Portella, M. Mendlowicz, S. Gleiser, E.S.F. Coutinho, I. Figueira,
Institutional h-index: The performance of a new metric in the evaluation of
Brazilian Psychiatric Post-graduation Programs, {\em Scientometrics}, {\bf 77} no. 2, 361-–368, (2008)}

\bibitem{Shibata03}{H. Shibata., Statistics of phase turbulence II,
{\em Physica A: Statistical Mechanics and its Applications}, {\bf 317} no. (3--4), 391--400, (2003).}

\bibitem{TsallisMA00}{C. Tsallis, and  M.P. de Albuquerque, Are citations of scientific papers a case of nonextensivity?,
{\em Eur. Phys. J. B}, {\bf 13}, 777--780, (2000).}

\bibitem{Tsallis99BJP}{C. Tsallis, Nonextensive Statistics:
Theoretical, Experimental and Computational
Evidences and Connections, {\em Brazilian Journal of Physics}, {\bf 29} no. 1, (1999)}

\bibitem{AlexandraTTMT04}{A.C. Tsallis, C. Tsallis, A.C.N. Magalhaes, and F.A. Tamarit,
Human and Computer Learning: An Experimental Study, {\em Complexus}, {\bf 1} no. 4, (2003).}

\bibitem{Tsallis88}{C. Tsallis, Possible Generalization of Boltzmann-Gibbs Statistics. {\em J.
Stat. Phys.}, {\bf 52}, pp. 479--487, (1988).}

\bibitem{Tsallis03}{C. Tsallis, C. Anteneodo, L. Borland, and R. Osorio., Nonextensive
statistical mechanics and economics, {\em Physica A: Statistical
Mechanics and its Applications}, {\bf 324} no. 1--2, 89--100,
(2003).}

\bibitem{Upadhyaya01}{A. Upadhyaya, J. Rieu, J.A. Glazier, and Y. Sawada, ``Anomalous
diffusion and non-Gaussian velocity distribution of Hydra cells in
cellular aggregates'', {\em Physica A: Statistical Mechanics and
its Applications}, {\bf 293} no. 3-4, 549-558, (2001)}

\bibitem{GellMannT04}
{M. Gell-Mann,  and C. Tsallis, eds., {\em Nonextensive
Entropy--Interdisciplinary Applications}, Oxford University Press,
New York, 2004.}

\bibitem{May97}{ R.M. May., The scientific wealth of nations. {\em Science}, {\bf 275}, 793--795, (1997).}

\bibitem{biblio} {An updated Bibliography is available at http://tsallis.cat.cbpf.br/biblio.htm, (2008).}


\end{thebibliography}
\end{document}